\documentclass[12pt]{article}

\def\slash#1{\setbox0=\hbox{$#1$}#1\hskip-\wd0\hbox to\wd0{\hss\sl/\/\hss}}
\usepackage{epsf, cite}
\usepackage{epsfig}                  
\usepackage[all]{xy}                 
\usepackage{amsfonts}
\usepackage{latexsym}

\usepackage{epsf, cite}

\makeatletter
\renewcommand\section{\@startsection {section}{1}{\z@}%
                                   {-3.5ex \@plus -1ex \@minus -.2ex}
                                   {2.3ex \@plus.2ex}%
                                   {\normalfont\large\bfseries}}
\renewcommand\subsection{\@startsection{subsection}{2}{\z@}%
                                     {-3.25ex\@plus -1ex \@minus -.2ex}%
                                     {1.5ex \@plus .2ex}%
                                     {\normalfont\bfseries}}
\makeatother

\def\lbldef#1#2{\expandafter\gdef\csname #1\endcsname {#2}}

\def\href#1#2{#2}

\def\beq{\begin{equation}}
\def\eeq{\end{equation}}
\def    \bea    {\begin{eqnarray}}
\def    \eea    {\end{eqnarray}}
\def \pl {\partial}
\def \KK {Ka\l u\.za-Klein}
\begin{document}
\pagestyle{plain}
\begin{titlepage}

\begin{center}

\hfill{QMUL-PH-2005-17} \\
\hfill{DAMTP-2005-115} \\

\vskip 1cm

{{\Large \bf M-theory and the string genus expansion}} \\

\vskip 1.25cm {David S. Berman\footnote{email: D.S.Berman@qmul.ac.uk}}
\\
{\vskip 0.2cm
Queen Mary College, University of London,\\
Department of Physics,\\
Mile End Road,\\
London, E1 4NS, England}\\
{\vskip 0.2cm
and}\\
{\vskip 0.2cm {Malcolm J. Perry\footnote{email: malcolm@damtp.cam.ac.uk}}}\\
{\vskip 0.2cm Department of Applied Mathematics and Theoretical Physics,\\
Centre for Mathematical Sciences,\\University of Cambridge,\\
Wilberforce Road,\\
Cambridge CB3 0WA,\\ England.}\\

\end{center}
\vskip 1 cm

\begin{abstract}
\baselineskip=18pt

The partition function of the membrane is investigated. In particular, the
case relevant to perturbative string theory of a membrane with topology
$S^1 \times \Sigma$ is examined. The coupling between the string world
sheet Euler character and the dilaton is shown to arise from a careful
treatment of the membrane partition function measure. This demonstrates 
that the M-theory origin of the dilaton coupling to the string world 
sheet is quantum in nature.

\end{abstract}

\end{titlepage}

\pagestyle{plain}

\baselineskip=19pt

\section{Introduction}

The basis for string perturbation theory is the coupling of the world
sheet Euler character, $\chi,$ to the dilaton, $\Phi$ ~\cite{polyakov}.
It is this
coupling that determines the relationship between the topology of the
string world sheet and the string coupling constant and allows string
theory at small coupling to be organised in terms of a genus
expansion. To see this, consider how the string coupling constant $g_s$ 
controls the amplitude
for a closed string to split into two. This is the so-called ``trousers''
diagram. To increase the genus of any amplitude
by one results in an extra multiplicative factor of $g_s^2$ for that
amplitude. An amplitude corresponding to a diagram with $e$ 
external legs and $l$ loops, will therefore  contain a factor of
$g_s^{e-2+2l}$.
Orientable two-dimensional surfaces are classified by their genus $g$
or Euler character $\chi$ together with the number of ends (corresponding
to external legs) where $\chi=2(1-g)$ and $g=l+{e/2}$. Thus each
amplitude is weighted by a factor of $g_s^{-\chi}.$ 

Despite the well known connection between membranes,
fundamental strings and D-branes
\cite{townsend,bergshoeff,witten,townsend2}{\footnote{There is interesting 
work on the quantum membrane and U-duality in string theory
\cite{boris} which looks into quantum aspects of the membrane and the
relation to string dualities.}}, the M-theory origin of the string world sheet
Euler character coupling to the dilaton has been regarded as something
of  a puzzle. 
To appreciate why the M-theory interpretation for this coupling has so
far remained mysterious, let us make a few observations that will 
be important in understanding the M-theory origin of this term. In
what follows, we will consider only string  world sheets without boundaries
ie. vacuum diagrams; the extension to include boundaries is
essentially trivial. 
Type IIA string theory
is obtained from M-theory by compactification on a \KK ~circle of radius 
$R_{11}$.
In string theory, the coupling of the dilaton to the world sheet $\Sigma$
is given
by a contribution to the string action of
\beq 
S_{\Phi}= {1 \over {4\pi}} \int_{\Sigma} d^2 \sigma \sqrt{\tilde\gamma}
{} R^{(2)}  \Phi 
\eeq
where $R^{(2)}$ is the Ricci scalar of the world-sheet metric $\tilde\gamma$.
Choosing a constant dilaton and using the Gauss-Bonnet theorem
\beq
{1 \over {4 \pi}} \int_{\Sigma} d^2 \sigma \sqrt{\tilde\gamma}  R^{(2)}
=\chi(\Sigma) \eeq
reduces the action  to 
\beq
S_{\Phi} = \chi(\Sigma)\Phi \, \, .
\eeq
Now, when  one makes a \KK ~reduction of
$11$-dimensional supergravity on a circle of radius $R_{11}$, down to
$10$-dimensional type IIA supergravity, one finds that the dilaton is
given by
\beq
e^\Phi =  \left( { R_{11} \over { l_p }
} \right) ^{3 \over 2} \, \, , \label{mstringrelations}
\eeq
Hence, in M-theory language, 
\beq 
g_s =  \left( { R_{11} \over { l_p }
} \right) ^{3 \over 2} \, \, ,
\eeq
and after choosing units such that $l_p=1$
\beq
S_{\Phi}={3 \over 2} \chi {\rm{log}}(R_{11}) \, \, \, .
\eeq
Both the absence of a factor of $\alpha '$ in the above action
and the logarithmic dependence on $R_{11}$ suggest that the origin of 
this term is not classical 
in nature.  It is more likely that such a term arises from a quantum 
effective action.
There is also the rather
trivial observation that one would like to be able to {\it{lift}} the
Euler character, $\chi(\Sigma)$ to three dimensions, so as to be able to give
a membrane interpretation. Finding the correct quantity in three
dimensions that when evaluated in $S^1 \times \Sigma$, where $\Sigma$
is a Riemann surface, is therefore part of the puzzle.

The approach that we will adopt is to describe the fundamental 
string as a membrane with world volume topology $S^1 \times \Sigma$ with 
$\Sigma$ being some Riemann surface. The membrane will be restricted to
wrap once around the M-theory circle such that the world
volume $S^1$ will be identified with the M-theory circle. We will
truncate to the zero mode sector of the circle. That is there will be
no dependence of any of the fields on circle direction, other than the
winding mode. This is
certainly justified if the M-theory circle is small since any
excitations will be very heavy. One can also interpret this as
isolating the pure fundamental string sector since any
dependence on the M-theory circle will be associated with D0 branes in
the string theory.

We will then describe the wrapped membrane 
partition function. In particular, we will be interested in
calculating the measure for a membrane with 
the topology relevant for the string, that is   $S^1 \times \Sigma$. We will 
then show that a careful treatment of this measure naturally gives rise 
to the correct dilaton coupling when interpreted from a string world 
sheet point of view.

\section{The membrane}

The bosonic part of the  action for the M-theory $2$-brane is, 
in Howe-Tucker form, 
\bea
S_{M2}[X,\gamma;G,C]&=& {T_{M2} \over 2} \int_{M^3} d^3\sigma \sqrt{\gamma} (
\gamma^{\mu \nu} \pl_{\mu} X^I \pl_\nu X^J G_{IJ} - 1 \nonumber \\
&+& \epsilon^{\mu  \nu \rho} \pl_\mu X^I \pl_\nu X^J \pl_\rho X^K C_{IJK} )
 \label{braneaction} \eea
where $T_{M2}$ is the tension of the M2-brane,  $\gamma_{\mu \nu}$  is 
the world-volume metric, $X^I(\sigma)$ specify the location of the brane 
in the target space. In what follows, upper case latin indices $I,J,K, \ldots$
are $11$-dimensional target space indices whereas greek indices are
world volume indices, and so $\sigma^\mu$ are the world volume coordinates.
$G_{IJ}$ and $C_{IJK}$ are respectively the background
metric and three form potential of eleven dimensional supergravity.
There are known obstacles for treating the membrane as one would the
string. For example, there is no discrete spectrum of states that
allows one to identify membrane states with space time excitations
\cite{nicolai}. It is also not obviously a renormalisable theory in
an arbitrary spacetime, even one that obeys the supergravity equations of
motion. Yet there are various consistency checks, mostly from the
supermembrane.  $\kappa$ symmetry of the supermembrane is consistent with
 the eleven
dimensional supergravity equations, \cite{townsend}. 
After world volume dualisation of one the scalar
fields, the membrane action can be identified with the D2 action
\cite{townsend2}. The relation of the BPS sector of the membrane to
dualities in string theory has been studied in \cite{boris} and important work on M-theory loops and higher derivative corrections to the string effective action corrections has been done in \cite{green}.

What we now
propose, following analogous work of Polyakov for the string, is to
take seriously the idea of a membrane partition function, $Z$ given by:
\beq
Z= \sum_{\rm topologies}
\int { { {\cal{D}} X {\cal{D}} \gamma } \over {{\rm{Vol(Diff_0)}}}}
e^{-S_{M2} [X,\gamma;G,C] } \, \, .
\label{m2partition} \eeq
or more precisely its supersymmetric extension. (For the present
we will ignore the Fermionic sector which appears to be irrelevant 
to our considerations regarding the origin of the dilaton coupling).
As with the string, the integrals are taken over all $X^I$
and world volume metrics, $\gamma_{\mu \nu}$ and then
summed over all topologies of the world volume.
Since the action is diffeomorphism invariant
one divides by the volume of three dimensional diffeomorphisms. 
(For the string it would be the product of diffeomorphisms and Weyl
transformations.)

To carry out the integral over world volume metrics, $\gamma$, one 
has to make an orthogonal decomposition of the deformations into those that are
pure diffeomorphisms and those which are physical. The decomposition
of the measure will then introduce a Jacobian which is the
Faddeev-Popov determinant. One is then free to
gauge fix and integrate over the pure diffeomorphism part of $\gamma$
leaving only the physical degrees of freedom, the {\it{moduli} } of
the three manifold. (The division by the
volume of the diffeomorphism group then cancels the integral over the
pure diffeomorphism part of $\gamma$.) 

Thus after gauge fixing and integrating over pure diffeomorphisms one
is left with the rather formal expression:
\beq
Z= \sum \int J d \{ {\rm{moduli}} \} {\cal{D}} X
e^{-S_{M2}[X,{\rm{moduli}};G,C]}  
\eeq
where $J$ denotes the Jacobian associated with the decomposition into
diffeomorphisms and moduli. Evaluating this for arbitrary membrane
topologies is certainly problematic since we do not have a
classification of three dimensional manifolds and their moduli. What
we will do instead is consider evaluating the above for a class of fixed
topologies where the moduli are known. The case
relevant to the fundamental string is where $M^3=S^1 \times \Sigma$ and 
the membrane is wrapped once around the M-theory circle, fixing the $S^1$
modulus. We will also truncate to the zero mode sector of the circle. 
The membrane moduli space is now isomorphic to the moduli space of 
Riemann surfaces.

First, let us choose a gauge where,
\beq
\sigma^3 = X^{11} \, , \qquad \gamma_{3\, 3}=1  \, 
\eeq
and since we have a trivial bundle we can also fix $\gamma_{i \,
  3}=0$. (We have also fixed the target space metric $G_{11,\,11}=1$.)
When the $S^1$ bundle is nontrivial we will not be able to make this
gauge choice and it is a important question how to extend this analysis
for non-trivial circle bundles.
After this partial gauge fixing the world volume metric is now:
\bea
 \gamma_{\mu \nu} = \left( \begin{array}{ccc}  
\tilde{\gamma}_{{i} {j}} &0  \\
            0 & 1   \end{array} \right) 
\label{reduction} \eea
where ${\tilde{\gamma}}_{{i} {j}}$ denotes the string world sheet
metric, $i,j$ are two dimensional world sheet indices.
The radius of the $S^1$ is fixed to be $R_{11}$ thus $\sigma^3 \in [0,R_{11}]$.
The action (\ref{braneaction}) then reduces to the usual action for the 
string, 
without the dilaton coupling term. 
\beq
S={1 \over {4 \pi \alpha' }} \int d^2 \sigma \sqrt{\tilde{\gamma}} (
\tilde{\gamma}^{i j} \pl_{i} X^I \pl_j X^J G_{IJ}  \\
+ \epsilon^{ij} \pl_i X^I \pl_j X^J B_{IJ})
\eeq
with $\alpha' = 1/2\pi T_{M2}R_{11}$.
Note, this action now has conformal symmetry. If one had taken a rather
more general \KK ~ansatz then we would have found
that the two-dimensional Weyl 
symmetry
was indeed part of the three-dimensional diffeomorphism symmetry, as
described in  \cite{Ana}.

Therefore the two dimensional conformal symmetry may be regarded as part of the three-dimensional 
diffeomorphism symmetry, \cite{Ana}.
Note,  if one were to
include the \KK ~ modes of the circle the conformal invariance
would be broken. 

Since the $S^1$ modulus of the membrane is fixed, one may now follow the usual
procedure of string theory to calculate the Jacobian, $J$ and evaluate 
the path integral for a given topology of the Riemann
surface. Following, \cite{d'hoker} 
the space of Riemann surface metrics is 
parameterised by ${\tilde{\gamma}}= exp(\delta v) e^{2
  \sigma} \hat{\gamma} $. Where $\hat{\gamma}$ is
transverse to the orbits of ${\rm{Weyl}}(\Sigma) \times
{\rm{\widetilde {Diff}}}_0(\Sigma)$
and $exp(\delta v)$ denotes a diffeomorphism generated by the vector
field $\delta v$. ${\rm{\widetilde{Diff}}_0}$ denotes two dimensional diffeomorphisms.

Infinitesimally, the joint action of Weyl transformations and 
diffeomorphisms on the
two dimensional metric is given by:
\beq
\delta \gamma_{ij} = (2 \delta \sigma + \nabla^k \delta v_k )
\gamma_{ij} + 2 (P_1 \delta v )_{ij}
\eeq
where:
\beq
P_1 (\delta v)_{ij}  = {1 \over 2} (\nabla_i \delta v_j +
\nabla_j \delta v_i
- \gamma_{ij} \nabla_k \delta v^k )\, \, .
\eeq
which maps vectors into tracefree symmetric rank two tensors.
We also introduce the adjoint map
\beq
P_1^\dagger (\gamma)_j=-2 \nabla^i \gamma_{ij} \, \, 
\eeq
from a symmetric tracefree two tensor to a vector.
It will be useful to treat the following three cases separately, genus
$g \geq 2$, the torus, $g=1$ and the sphere, $g=0$. The case
where $g \geq 2$ is the simplest because there are no conformal
Killing vectors, ie. the kernel of $P_1$ is null. The dimension of
the moduli space is given by the kernel of $P_1^{\dagger} $ 
which for $g \geq 2$ is given by:
\beq
{\rm{Ker(P_1^\dagger)}} = 6g-6 = - 3 \, \, \chi \, .
\eeq
The Jacobian $J$ is given by:
\beq
J= {{\rm{det}} (P_1^\dagger P_1)}^{1\over2} \, .
\eeq
(Note, for the case $g \geq 2$ there are no zero modes of $P_1$). 
This leads to the partition function:
\beq
Z= \int {\cal{D}} X  \, d^{6g-6} m \, \, {{ {\rm{det}} 
(P_1^\dagger P_1)}}^{1\over2} \, {{dv\,d\sigma} \over
 {{\rm{Vol(\widetilde{Diff}}_0 \times {\rm Weyl)}}}} 
\,  e^{-S[X,m;G,B]}
\label{m2partition2}
\eeq
with $d^{6g-6} m$, denoting the integration over the $6g-6$ dimensional
moduli space of the Riemann surfaces with genus g. The integration
over $v$ and $\sigma$ will then cancel the factor of
${\rm Vol({\widetilde{Diff}}_0}\times {\rm Weyl)}$ in 
the denominator.  
One now calculates the measure on moduli space, being especially
careful with the normalisation, \cite{d'hoker, alvarez}.
So far the above has been exactly as for the string. However, the
normalisation of the string and wrapped membrane moduli will be
different due to the presence of $S^1$. 

To calculate the measure on moduli space one typically writes the
moduli in terms of quadratic differentials (see
\cite{d'hoker,alvarez}). Then the norm of each quadratic differential
is given by:
\beq
||\gamma_{\mu \nu}||^2=\int d^3 \sigma \sqrt{\gamma} (\gamma^{\mu \rho}
\gamma^{\nu \sigma} -c \gamma^{\mu \nu} \gamma^{\rho\sigma}) \delta \gamma_{\mu \nu} \delta \gamma_{\rho
  \sigma} \, \, . \label{norm}
\eeq
(c is a constant upon which nothing physical will depend and so
is set to zero from now on). The measure on moduli space is then given
by the product of norms
(assuming an orthogonal basis for quadratic differentials).
We now take the above expression and calculate the norm of the moduli
space of the wrapped membrane in terms of the moduli space of the
string. Evaluating the norm (\ref{norm}) with the metric (\ref{reduction}) and
integrating over the $S^1$ gives the relation between the norm of a quadratic differential for
the wrapped membrane and that of the string:
\beq 
||\delta \gamma_{\mu \nu}||= \sqrt{R_{11}} ||\delta \tilde{\gamma}_{ij}||
\label{scaling} \eeq
Thus, the relation between the wrapped membrane
moduli space measure $d^{6g-6}m$ and string moduli space measure $d^{6g-6}\tilde{m}$ is given by:
\beq
d^{6g-6}m = \left( R_{11} \right)^{3g-3} d^{6g-6} \tilde{m} \, . 
 \label{scaling2}
\eeq
Inserting this into (\ref{m2partition2}) we write the wrapped membrane
partition function in terms of string normalised moduli as follows:
\beq
Z= \int {\cal{D}} X \,  \left(R_{11}\right)^{3g-3}
\, d^{6g-6} {\tilde{m}} \, \, 
{{ {\rm{det}} (P_1^\dagger P_1)}}^{1 \over2} \, \, e^{-S[X;G,B]} \, .
\eeq
We may then use the M-theory, string theory relations (\ref{mstringrelations})
 to write
the $R_{11}$ factors arising from the wrapped membrane moduli space
measure in terms of the dilaton, $\phi$ and the string world sheet
Euler character, $\chi$.
This gives,
\beq
Z=   \int {\cal{D}} X  \, d^{6g-6} {\tilde{m}}  \,  {{{\rm{det}}
  (P_1^\dagger P_1)}}^{1\over2} \, \,  e^{-S[X;G,B]-\phi \chi}  \, \, .
\eeq
So remarkably the partition function of the wrapped membrane
reproduces that of the string including the dilaton coupling to the
world sheet Euler character. The latter is essential an effective
action coming from a proper treatment of the wrapped membrane moduli space 
measure.

To complete the discussion we now need to consider the case of the
sphere and the torus where $P_1$ has zero modes (ie. there are
conformal Killing vectors). Again we will follow
\cite{d'hoker,alvarez} treatment of the string.

The measure must now be modified so as not to integrate over zero modes:
\beq
{\cal{D}} \gamma_{\mu \nu} = ({\rm{det}}'P_1^\dagger P_1)^{1 \over 2}
d{} 'v^\mu d\sigma \,  d{\rm{moduli}}
\eeq
where the prime denotes the restriction to the space orthogonal to the
Kernel of $P_1$ ie. $(Ker(P_1))^\perp$. This may then be written as:
\beq
{\cal{D}} \gamma_{\mu \nu} = {1 \over {\rm{Vol(KerP_1)}}} 
({\rm{det}}'P_1^\dagger P_1)^{1 \over 2}
dv^\mu d\sigma \, d{\rm{moduli}}
\eeq
where all reparameterisations are now included in $d v^\mu$. We
then must see how the norm of the conformal Killing vectors scales
between the wrapped membrane and the string just as for the norms of
the quadratic differentials ie. the moduli. Again using, (\ref{norm}) one sees that the relation
(\ref{scaling}) is also followed by the norm of the conformal Killing vectors. 
Thus,
\beq
{\rm{Vol(KerP_1)}} = (R_{11}) ^{{1 \over 2} \rm{Dim(Ker(P_1))}}
  {\rm{Vol(Ker{\tilde{P}}_1)}}  \, \, ,
\label{scaling3}
\eeq
where the tilde denotes $P_1$ in string variables. Combining the above
contributions of both the rescaling of the moduli space measure
(\ref{scaling2}) and
the rescaling of the volume of the kernel of $P_1$ ie. the volume of
conformal Killing vectors (\ref{scaling3}) one obtains the overall
scaling between the wrapped membrane partition function measure and the string
partition function measure to be:
\beq
\int { { {\cal{D}} \gamma } \over {\rm{Vol(Diff_0)} }  } =\int
  { { {\cal{D}} {\tilde{\gamma}}} \over {\rm{Vol(\widetilde{Diff}}_0}
 \times {\rm Weyl)} } 
(R_{11})^{ {1 \over 2} ( {\rm{DimKerP}}_1^\dagger - {\rm{DimKerP}}_1  ) }
\, \, ,
\eeq 
where we have used that the dimension of moduli space is given by $\rm{DimKerP_1^\dagger}$.
We may now invoke the Riemann-Roch theorem:
\beq
 {\rm{DimKerP}}_1^\dagger - {\rm{DimKerP}}_1 =-3 \, \chi
\eeq
to write the total rescaling as:
\beq
{{\cal{D} \gamma} \over {\rm{Vol(Diff_0)}}} = 
\, { { {\cal{D}} {\tilde{\gamma}} } \over
  {\rm{Vol(\widetilde{Diff}}_0 \times
{\rm Weyl)}} }
(R_{11})^{-{3 \over 2} \chi } =  
{ { {\cal{D}} {\tilde{\gamma}} } \over {\rm{Vol( {\widetilde{Diff}}_0 }
  \times {\rm Weyl)}} }    \, 
e^{- \phi \chi} \, .
\eeq
Thus we have again produced, as required the dilaton coupling to the
world sheet Euler characteristic from the rescaling of the membrane measure.

\section{Discussion}

The first question to address is, given what we know
about M-theory, to what extent was the above observation inevitable. The usual
relations between M-theory and string theory are made in the classical
sector and then (often using BPS type arguments) extrapolated into
the quantum regime where we know little about M-theory. The key
relation between M-theory and string theory is between the string
coupling and the radius of the eleventh dimension, $g_s=(R_{11})^{3
  \over 2}$. This is usually derived from the identification
of the classical action of the wrapped membrane with the fundamental
string and the momentum around the M-theory circle with the D0 brane.

From the point of view presented here, that relation (in
particular the power of ${3 \over 2}$) is a
consequence of two facts. First, the dimension of moduli space of Riemann surfaces
scales like $-3\chi$ and secondly, the measure of the moduli space of $S^1
\times \Sigma$ is related to the measure of the moduli space of a
Riemann surface, $\Sigma$ by a factor of $(R_{11})^{ {1 \over 2}
  \rm{dim(moduli space)}}$. This then gives the appropriate coupling
  between the dilaton and the Euler characteristic of the
  string word sheet in M-theory language. It did not have to be this
  way. The M-theory, string theory identification could have been
  anomalous for the membrane, that
  is, although the classical actions are equivalent; the partition functions
  need not be. What we have shown here is that the measure of the
  wrapped membrane partition function does indeed match the string
  theory partition function measure. 

Importantly, the string coupling to the
  world sheet Euler character is not put in by hand but arises naturally from
  the membrane partition function measure as one would hope for in a
  nonperturbative theory. 

One might also ask, what is the justification for our ad hoc
restriction of the membrane topology and the truncation to the zero
modes of $\pl_{\theta}$, ie. none of the fields have dependence on the
M-theory circle.

Consider the expansion of the world volume fields on a circle, 
\bea
X^{11}&=& N \theta  + \sum_k (e^{ik \theta} + e^{-ik \theta})
+ p \tau \\
X^i&=& \sum_l (e^{i l \theta} + e^{-il\theta})  \, \, .
\eea

N is the membrane winding number and corresponds to the number of
fundamental strings. $p$ is the momentum around the M-theory circle and
corresponds to the D0 brane charge. In order to satisfy the boundary
conditions $k$ and $l$ must be integer. $k$ corresponds to the
$D0-{\bar{D}}0$
states while $l$ corresponds to the states of  the non wrapped
membrane. We can ignore these modes if the M-theory circle is small
with respect to the Planck scale. That is weak coupling in string theory 
language. 
The action for the non wrapped membrane will be larger than for the
wrapped membrane and so the wrapped membrane will dominate. The D0
brane states will also be more massive than the states we are
considering. Thus we truncate to the $k=0, l=0, p=0, N=1$ sector. 
Essentially leaving a single fundamental string. This is of course
consistent with what we know about string theory in that the string is
a consistent object at weak coupling but at strong coupling the
contribution of nonperturbative objects such as D-branes becomes important.
Redoing the above calculation of the membrane measure but including these 
truncated states may give interesting
couplings between the dilaton and D-branes states bound to the fundamental 
string.

Lastly, we ignored the possibility of the world volume having boundaries, 
as would be exzpected when there are external states attached to the 
string worldsheet. All of the results above can be trivially  extended 
to cover that possibility.

Integrating over metrics modulo diffeomorphisms, is obviously a topological
invariant. It would be interesting to know if this can be evaluated in 
other circumstances,
ie. for membranes with more nontrivial three dimensional
topologies. This would then give more credence  to membrane instantons.

\section{Acknowledgements}

We wish to thank the following people for discussions: James Bedford, 
Jan de Boer, Nick Dorey, Michael Green, Sean Hartnoll,
Jim Liu, Lubos Motl, Is Singer, Paul Townsend, Neil Turok and in
particular Robert Helling for initial discussions. DSB is supported
by EPSRC grant GR/R75373/02 and would like to thank DAMTP and 
Clare Hall 
Cambridge for continued support. This work was in part 
supported by the EC Marie Curie
Research Training Network, MRTN-CT-2004-512194.

\end{document}